\documentclass[singlecolumn, switch]{article}

\usepackage{preprint}

\usepackage[utf8]{inputenc} 
\usepackage[T1]{fontenc}    
\usepackage{hyperref}
\usepackage{url}            
\usepackage{booktabs}       
\usepackage{bm}
\usepackage{amsfonts}       
\usepackage{amsmath}
\usepackage{amssymb}
\usepackage{amsthm}
\usepackage{nicefrac}       
\usepackage{microtype}
\usepackage{cleveref}
\usepackage{mathtools}
\usepackage{romannum}
\usepackage{authblk}
\usepackage{xcolor}
\usepackage{algpseudocode}
\usepackage{algorithm}
\usepackage{color,soul}
\usepackage{caption}
\usepackage{lipsum}
\usepackage{fancyhdr}       
\usepackage{graphicx}       
\graphicspath{{media/}}     
\usepackage{enumitem}
\usepackage{lineno}
\makeatletter
\newcommand{\printfnsymbol}[1]{%
  \textsuperscript{\@fnsymbol{#1}}%
}

\NewDocumentCommand{\ARef}{ s s m }{%
    \IfBooleanTF{#2}{}{%
        \cref{#3}%
    }%
    \IfBooleanT{#1}{%
        \IfBooleanF{#2}{%
            , %
        }%
        line~\ref{#3}%
    }%
}

\newtheorem{theorem}{Theorem}

\makeatother

\title{TOBACO: Topology Optimization via Band-limited Coordinate Networks for Compositionally Graded Alloys}

\author[1]{Aaditya Chandrasekhar}
\author[1]{Stefan Knapik}
\author[1]{Deepak Sharma}
\author[2]{John Reidy}
\author[2]{Ian McCue}
\author[1]{Jian Cao}
\author[1,*]{Wei Chen}

\affil[1]{Department of Mechanical Engineering, Northwestern University, Evanston, IL, USA
\authorcr
   \{\tt aadityacs, deepak.sharma, jcao, weichen\}@northwestern.edu, stefan.knapik@u.northwestern.edu}

\affil[2]{Department of Materials Science and Engineering, Northwestern University, IL, USA
\authorcr \tt
   ian.mccue@northwestern.edu, johnreidy2028@u.northwestern.edu}
\affil[*]{Corresponding author}

\begin{document}
\maketitle

\begin{abstract}

Compositionally Graded Alloys (CGAs) offer unprecedented design flexibility by enabling spatial variations in composition; tailoring material properties to local loading conditions. This flexibility leads to components that are stronger, lighter, and more cost-effective than traditional monolithic counterparts. The fabrication of CGAs have become increasingly feasible owing to recent advancements in additive manufacturing (AM),  particularly in multi-material printing and improved precision in material deposition. However, AM of CGAs requires imposition of manufacturing constraints; in particular limits on the maximum spatial gradation of composition.

This paper introduces a topology optimization (TO) based framework for designing optimized CGA components with controlled compositional gradation. In particular, we  represent the constrained composition distribution using a band-limited coordinate neural network. By regulating the network’s bandwidth, we ensure implicit compliance with gradation limits, eliminating the need for explicit constraints. The proposed approach also benefits from the inherent advantages of TO using coordinate networks, including mesh independence, high-resolution design extraction, and end-to-end differentiability. The effectiveness of our framework is demonstrated through various elastic and thermo-elastic TO examples.

\end{abstract}


 \vspace{1cm}
 

\section{Introduction}
\label{sec:intro}

Topology optimization (TO) \cite{sigmund2013topology} is a class of computational methods that optimizes material distribution within a given design domain, subject to specified objectives and constraints. These methods can navigate complex design landscapes that challenge conventional, intuition-based approaches. However, conventional TO focused on designing with a single material may be insufficient when multiple, conflicting functional requirements are present \cite{magerramova2016TurbineBladeTO}. For instance, a turbine blade requires varied local properties such as surface oxidation resistance, tip creep resistance, and low thermal expansion at its base, while minimizing weight and reducing the amount of critical materials \cite{mondal2021thermalBarrierCoating}. It is difficult for a single material to satisfy all these demands.  To balance these conflicts, researchers have explored designs incorporating multiple materials \cite{xia2008FGMs, dunning2015FGMLevelSet, conlan2018FGMCompliantMech}; in particular compositionally graded alloys (CGAs) \cite{liu2018ReviewTOforAM}.

\begin{figure}[h]
 	\begin{center}
		\includegraphics[scale=0.65,trim={0 0 0 0},clip]{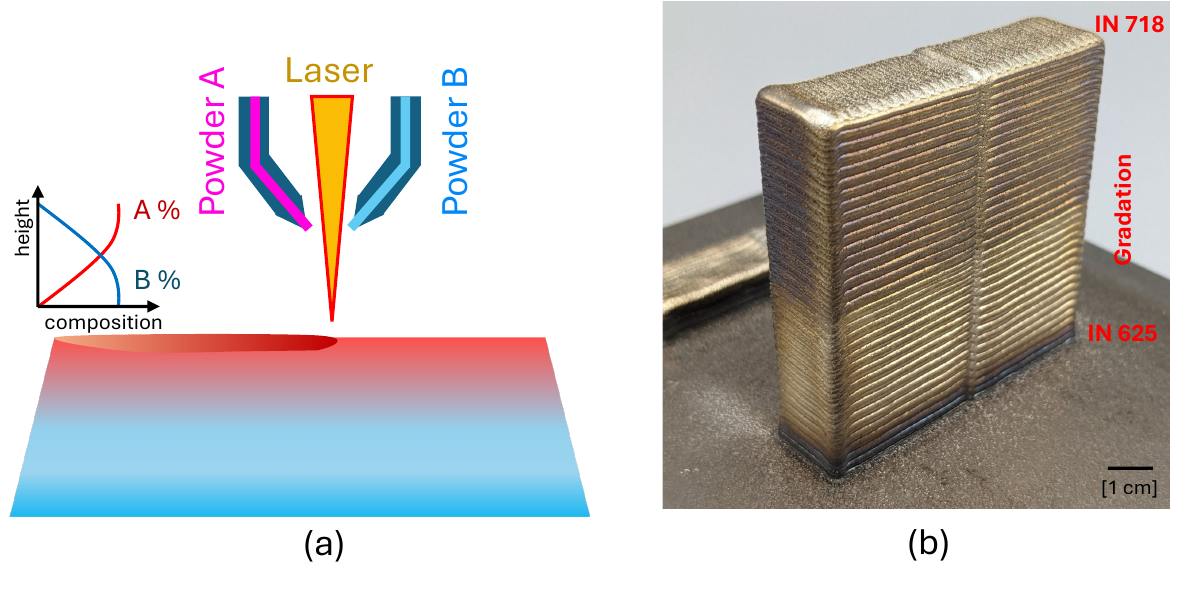}
 		\caption{(a) Illustration of Additive Manufacturing of CGA using Directed Energy Deposition \cite{zha2024situ}. (b) CGA with the bottom layer printed with $100\%$ IN625, top layer with  $100\%$ IN718 and graded in between. Image courtesy of  Rujing Zha and Rowan Rolark. }
 		\label{fig:FGM_print}
	\end{center}
 \end{figure}

CGAs (\Cref{fig:FGM_print}) provide enhanced design flexibility by allowing material composition to vary spatially \cite{chan2021metaset, chan2022remixing}. This enables the precise tailoring of material properties to local requirements, offering a way to balance conflicting design objectives \cite{li2020reviewFGM,price2025fly}. However, this flexibility complicates both the design and manufacturing processes. Recent advances in Additive Manufacturing (AM) have enabled the fabrication of complex metal \cite{singh2021FGMDED, salcedo2018FGMMJ} and polymer multi-material parts, including those with graded compositions. Despite this progress, the AM of multi-material components \cite{careri2023AMHX, pinelli2022AMTurbine, gradl2018AMRocketEngine} imposes its own constraints, particularly limitations on the achievable composition gradation.

This paper focuses on incorporating these gradation constraints into the TO framework \cite{vatanabe2016ManufConsTO}. Such constraints arise from two primary sources. First, sharp transitions between dissimilar alloys can induce thermal stresses, leading to delamination \cite{chen2017FabCGA, su2020influenceCompGrad}. Second, manufacturing process parameters and machine capabilities inherently limit how rapidly the composition can be varied \cite{yang2023designFGM}.

\begin{figure}[h]
 	\begin{center}
		\includegraphics[scale=0.65,trim={0 0 0 0},clip]{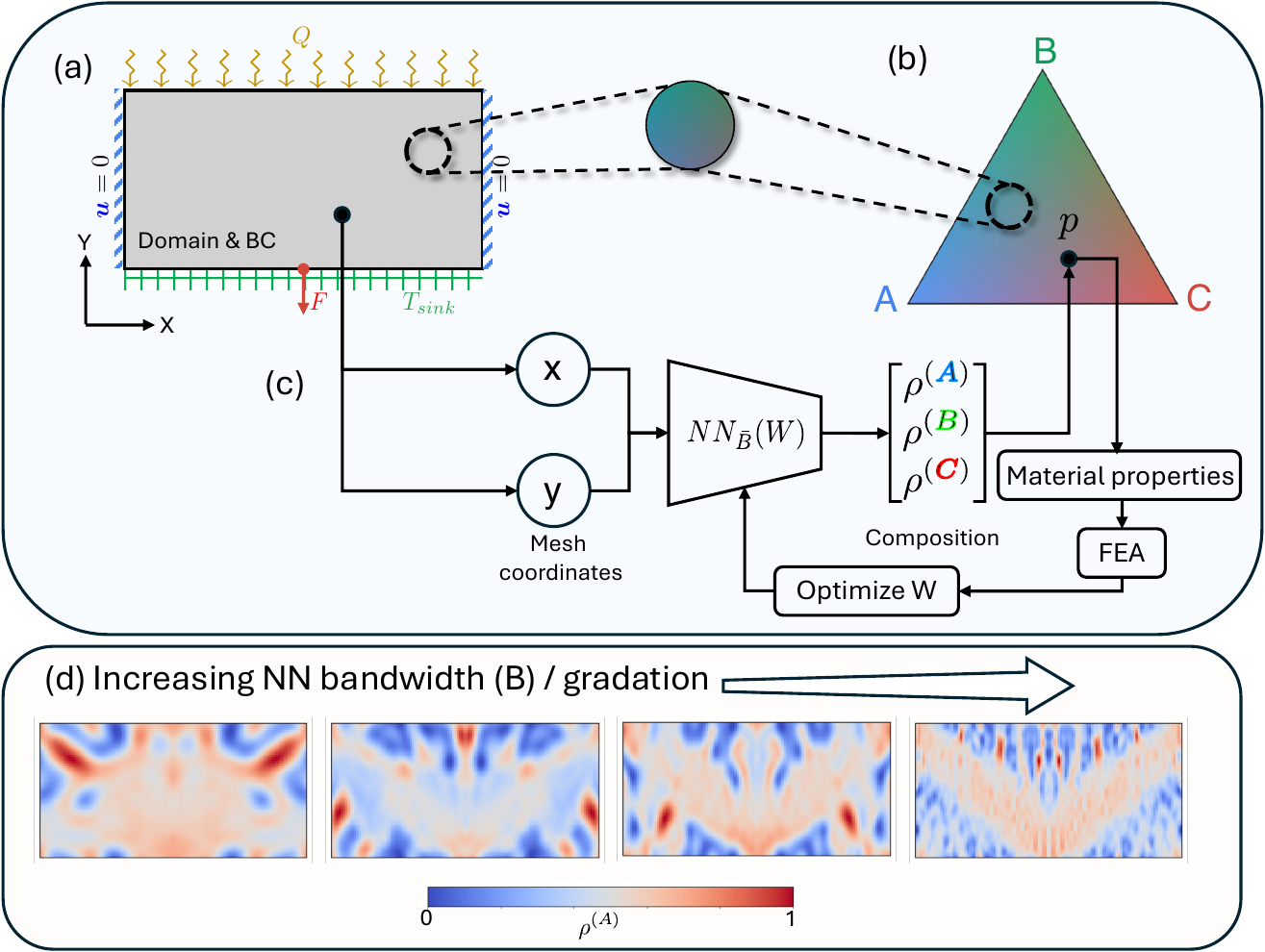}
 		\caption{Overview of the proposed method. The inputs are (a) a domain subjected to thermo-elastic loads and (b) a material system with composition and phase dependent properties. A neural network then maps coordinates to composition and material properties (c). The weights of the network are optimized to determine optimal material distribution. The gradation of the design composition is controlled by varying the coordinate network bandwidth, $\bar{B}$, to produce (d) optimized designs with varying gradation limits.}

 		\label{fig:graphical_abstract}
	\end{center}
 \end{figure}

To address these limitations, we introduce a method for imposing gradation constraints in the TO of CGAs. Our approach first formulates the gradation constraint within the Fourier space. We leverage signal processing theory \cite{lapidoth2017foundationInDigitalComm} to demonstrate that a constraint on the maximum gradation of composition directly corresponds to imposing a band limit on its Fourier spectrum. Then, extending a recently proposed neural network (NN) based design representation \cite{chandrasekhar2021tounn}, we represent the compositions using a band-limited NN \cite{lindell2022bacon} to implicitly enforce this maximum gradation constraint. Finally, we formulate a TO framework for designing structures composed of CGAs using this band-limited representation (\Cref{fig:graphical_abstract}). While our primary focus is on elastic and thermo-elastic problems, the proposed method is broadly applicable.

\section{Related Work}
\label{sec:relatedWork}

This work contributes to the expanding research aimed at combining the design flexibility of additive manufacturing (AM) with the computational design capabilities of topology optimization (TO) to develop higher-performing components \cite{cucinotta2019TO_AM_biomed, jihong2021reviewTO_AM_cons, zegard2016bridging_TO_AM}. Specifically, this study focuses on enhancing the manufacturability of CGA designs optimized via TO by explicitly incorporating manufacturing constraints.

Topology optimization encompasses various methods, such as density-based approaches \cite{sigmund2013_TO_review}, level-set approaches \cite{wang2003levelSet_TO} and geometric projections \cite{norato2015geometryProjTO}. Our research utilizes the density-based approach as its foundational framework. Density-based TO discretizes the design domain into finite elements and optimizes the fictitious material density in each element, traditionally used for single-material optimization. Recently, there has been significant interest in expanding TO methodologies to handle multi-material scenarios \cite{zuo2017multiMaterialTO}. These include designs featuring discrete domains of distinct materials \cite{gao2016MultiMaterialThermoElasticTO} and functionally graded materials (FGMs) characterized by smooth transitions between materials \cite{taheri2014thermoElasticFGMIsoGeomTO}. Such methods substantially enhance the design space by enabling spatial tailoring of material properties.

Functionally graded structures, including porous configurations, have found notable applications in aerospace and biomedical industries due to their superior structural performance and multiphysics capabilities \cite{klippstein2018additive, wu2022controllable, padhy2025voroto, hollister2005porous}. A prime aerospace example is turbine blades, which face simultaneous mechanical loads and thermal gradients, thus benefiting from spatially varying properties. Recent studies demonstrated that graded lattice structures optimized via TO significantly outperform homogeneous designs, enabling denser architectures near heated regions for improved heat conduction and lighter structures elsewhere to enhance cooling efficiency \cite{alkebsi2021design}. Moreover, FGMs have also proven valuable in thermal barrier coatings \cite{mondal2021thermalBarrierCoating}. Extensive reviews of multi-material components and FGM applications can be found in \cite{saleh202030FGMApplications, nazir2023multiMaterialAMReview}. CGA, a specialized category within FGMs , are increasingly utilized in critical aerospace, nuclear, and biomedical components to meet specific local requirements, such as corrosion resistance in one region and creep resistance in another, often necessitating the combination of distinct alloys \cite{wen2021LBF_CGA, kirk2021CGA,allen2024graph,tonyali2024additively}.

Manufacturing optimized CGA designs requires addressing specific process-related constraints. While manufacturing constraints for conventional subtractive processes, such as turning, milling,  drilling, have been widely explored \cite{vatanabe2016topology}; additive manufacturing techniques, such as powder-bed fusion or extrusion-based 3D printing, permit greater geometric freedom but introduce new constraints \cite{liu2018ReviewTOforAM}. Notably, overhang constraints are critical in AM, as unsupported overhangs beyond certain angles may cause fabrication failures \cite{ebeling2021topology}. Topology optimization methods typically handle these constraints by penalizing or explicitly restricting overhang features \cite{langelaar2016TO_supportStructure}. Another vital AM constraint is length-scale control, approached either exactly or approximately. Exact methods, while precise, involve complex implementations \cite{zhang2014explicit}. Approximate methods, including projection filters and morphological constraints \cite{zhou2015minimumLengthScaleFourier}, offer practical alternatives by indirectly enforcing feature-size limits \cite{sigmund1997design, chandrasekhar2022LengthScale}. For a detailed overview of AM constraints, readers are referred to \cite{ebeling2021topology}.

Crucially, gradation control is an essential constraint for CGAs, as overly abrupt compositional transitions can be non-manufacturable, cause material incompatibilities, or cause high thermal stresses \cite{zuback2019AM_FGM_ferrous}. Gradation constraints ensure smooth transitions by limiting the rate of compositional change across adjacent regions. Manufacturing processes impose explicit limits on gradation scales; for example, graded Al-W components have been manufactured with gradation that span 0–55 vol$\%$ W in Al over length scales of approximately 0.6–1.2 mm \cite{kelly2021DED_grad_limit}. In Directed Energy Deposition (DED), Ti6Al4V/Invar FGAs showed gradation increments around 3 vol$\%$ per layer across 32 layers \cite{yoo2023DED_gradlimit}. Laser powder bed fusion offers finer control at micron scales (0.02–0.1 mm) , while wire arc additive manufacturing (WAAM) typically achieves coarser gradation scales (0.5–1 cm) \cite{pandiyan2024PBF_gradLimit} due to larger bead sizes. Consequently, incorporating gradation constraints into the optimization framework is imperative for producing feasible and manufacturable designs.

In conventional element-based TO, such constraints are typically imposed by restricting the difference in composition fractions between adjacent elements \cite{almeida2010layout, silva2025topology}. However, this approach presents several limitations. First, calculating spatial gradients with finite differences on a discretized field can be inaccurate. Second, a discrete representation is fundamentally limited in its ability to model the continuous compositional variations inherent to CGAs \cite{allen2024CGAGraph}. Finally, the design resolution is coupled to the analysis mesh, which restricts the achievable design complexity. Overcoming these challenges necessitates a design representation that is continuous, mesh-independent, and allows for precise gradient control. Towards this, we employ a band-limited neural network (NN) to impose gradation constraints effectively.

The choice of a neural network for this task aligns with recent advancements in the field of design representation \cite{park2019deepsdf,mildenhall2021nerf}. Integrating machine learning \cite{woldseth2022TO_ML_review}, particularly neural networks, into topology optimization is gaining popularity for enhanced design representation and computational efficiency \cite{chandrasekhar2021tounn, sosnovik2019neural}. Furthermore, this paper primarily investigates neural network-based representation methods suitable for CGAs. Coordinate-based deep neural networks offer a continuous parametrization by mapping spatial coordinates directly to material density values, thus replacing discrete element-based densities \cite{chandrasekhar2021tounn}. Such networks have been extended successfully to multi-material TO applications \cite{chandrasekhar2021multi}. Additional approaches leveraging neural networks include representing material properties \cite{chandrasekhar2022integrating}, shapes \cite{padhy2024photos}, and microstructures \cite{padhy2024tomas, chan2021metaset}.

Recently, representation in Fourier space \cite{sitzmann2020SIREN, white2018FourierTO} (frequency domain) has emerged as a promising approach, describing topologies through truncated Fourier series whose coefficients are optimized. This Fourier-based method can be used to limit spectral bandwidth \cite{lindell2022bacon}. Building upon this concept, our work leverages Fourier-based neural networks to represent CGAs, inherently ensuring smooth and manufacturable gradation transitions.

\subsection{Contributions}
\label{sec:relatedWork_contributions}

This paper presents a TO framework for designing components with CGA using AM, specifically incorporating gradation manufacturing constraints. AM methods face limitations regarding the maximum allowable spatial gradation, primarily to limit thermal stresses and inherent process capabilities. In particular, we recognize that standard element-based TO methods struggle to model the continuous compositional variations inherent in CGAs. Moreover, discretization makes it challenging to accurately calculate and enforce spatial gradient constraints. Additionally, design resolution is inherently tied to the analysis mesh size, restricting achievable design complexity. To address these limitations, this paper introduces several key contributions:

\begin{enumerate}

\item \textbf{Band-Limited Neural Network Representation}: We employ a continuous, mesh-independent design representation using a band-limited coordinate neural network that maps spatial coordinates directly to material composition (\Cref{fig:graphical_abstract}(c)).

\item \textbf{Implicit Gradation Control}: We formulate the gradation constraint in the Fourier domain and leverage Bernstein's Inequality. This approach implicitly enforces the maximum spatial gradation limit by controlling the network's bandwidth, thereby eliminating the need for explicit constraints (\Cref{fig:graphical_abstract}(d)).

\item \textbf{End-to-End Differentiable Framework}: The proposed framework is fully differentiable, utilizing automatic differentiation to compute the gradients required for optimization. This streamlines the optimization process and enhances computational efficiency.

\end{enumerate}

The framework's effectiveness is demonstrated through several numerical experiments. First, we validate our central hypothesis using a 2D structural compliance minimization problem under thermo-elastic loads and a ternary material system (\Cref{fig:graphical_abstract}a-b). These results confirm that the network's bandwidth effectively controls the final design gradation. We then demonstrate the framework's ability to handle diverse manufacturing constraints by applying it to a thermo-elastic problem with varying isotropic and anisotropic gradation limits. Finally, a 3D thermo-mechanical optimization of a turbine blade confirms the method's practical utility for complex, real-world geometries.

\section{Proposed Method}
\label{sec:method}

\subsection{Overview}
\label{sec:method_overview}

\begin{figure}[h]
 	\begin{center}
		\includegraphics[scale=0.65,trim={0 0 0 0},clip]{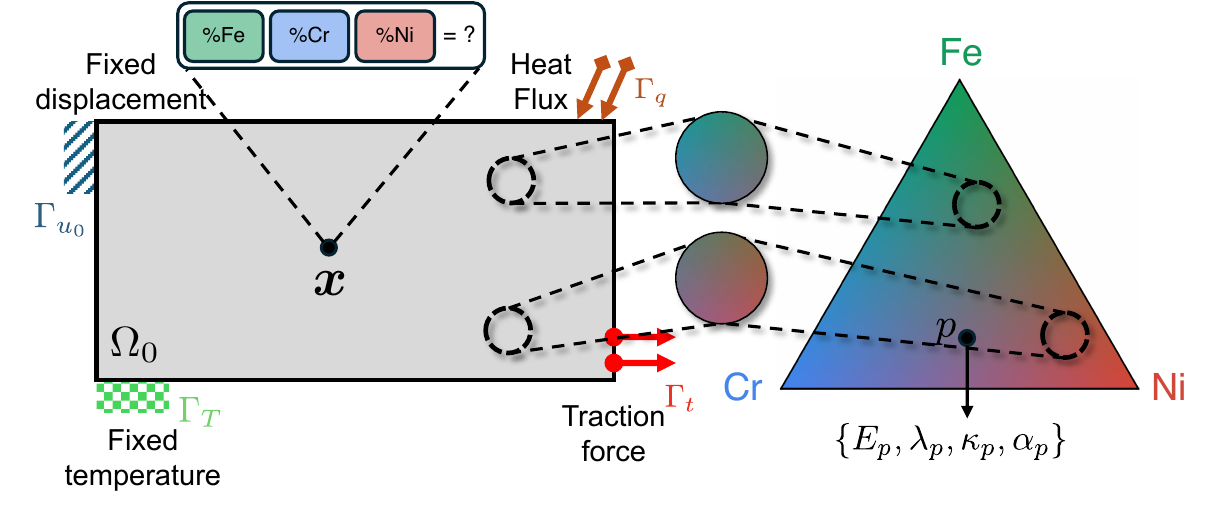}
 		\caption{The setup of the problem. Given a domain with loading and a material system (for e.g., Fe-Cr-Ni), determine the optimal composition at each point.}
 		\label{fig:problem_statement}
	\end{center}
 \end{figure}

In this study, we focus on the optimal spatial distribution of CGAs. We begin by assuming a domain $(\Omega_0)$ subjected to prescribed loads (\Cref{fig:problem_statement}) is prescribed. Furthermore, we assume a material system (such as Fe-Cr-Ni) wherein the fraction of constituent materials can be continuously varied is also prescribed. Additionally, we assume that material properties, such as Young's modulus $(E)$, thermal conductivity $(\kappa)$, mass density $(\lambda)$, and thermal expansion coefficient $(\alpha)$, can be determined for any given composition \cite{liu2010calphad_modulus,gheribi2012calphad_cond,campbell2024calphad_review}. Our objective is to compute an optimal design by determining the ideal material composition at each spatial location within $\Omega_0$. Notably, we require the distribution to satisfy imposed restrictions on the maximum spatial gradation of the composition.

Let us now consider the  formulation of the optimization problem. We consider a material system composed of $S$ components. The composition at any point $\bm{x} \in \Omega_0$ is described by the composition vector $\vec{\rho}(\bm{x}) = \{ \rho^{(1)}, \rho^{(2)}, \ldots, \rho^{(S)}\}$. Each component fraction, $\rho^{(j)}$, is bounded such that $\rho^{(j)} \in [0, 1]$. The component fractions must also satisfy the partition of unity constraint, $\sum\limits_{j=1}^S \rho^{(j)} = 1$. Furthermore, we impose a maximum gradation limit, $L_{max}$. This restricts the spatial gradient of the component fractions: $\left|\frac{\partial \rho^{(j)}}{\partial \bm{x}}\right| \leq \frac{1}{L_{max}}$.

We can then express a canonical optimization problem as:

\begin{subequations}
  \label{eq:optimizationIntroEquations}
  \begin{align}
    & \underset{\vec{\bm{\rho}} \; (\forall \bm{x} \in \Omega_0)} {\text{minimize}}
    & & J (\vec{\bm{\rho}}, \bm{u})
    \label{eq:opt_intro_objective} \\
    & \text{such that}
    & & \bm{R}(\vec{\bm{\rho}}, \bm{u}) = \bm{0}
    \label{eq:opt_intro_govnEqThermo} \\
    & & & g_i (\vec{\bm{\rho}}, \bm{u}) \leq 0 \; , \; i=1,\ldots, n
    \label{eq:opt_intro_massCons} \\
    & & & 0 \leq \rho^{(j)} \leq 1 \quad , \; j = 1, 2, \ldots, S
    \label{eq:opt_intro_boundCons} \\
    & & & \sum\limits_{i=1}^{S} \rho^{(i)}(\bm{x}) = 1 \; , \; \forall \bm{x} \in \Omega_0
    \label{eq:opt_intro_partitionUnityCons} \\
    & & & \left|\frac{\partial \rho^{(j)}}{\partial \bm{x}}\right| \leq \frac{1}{L_{max}} \quad , \; j = 1, 2, \ldots, S \;, \; \forall \bm{x} \in \Omega_0
    \label{eq:opt_intro_maxGradationCons}
  \end{align}
  \label{eq:optimization_intro}
\end{subequations}

Here, $J$ is the objective metric to be minimized, such as structural compliance. $\bm{R}$ represents the residual from the governing physical equations. The functions $g_i$ denote additional inequality constraints, for instance, a mass constraint. The bound constraints, partition of unity, and maximum gradation constraints are expressed in \Cref{eq:opt_intro_boundCons,eq:opt_intro_partitionUnityCons,eq:opt_intro_maxGradationCons}, respectively.

\subsection{Design Representation}
\label{sec:method_bandLimitedRepresentation}
Typically, the optimization problem in \Cref{eq:optimization_intro} is solved using density based TO methods \cite{sigmund2013topology} where the design variables are discretized element densities (component fractions). However, this discretized design representation presents challenges for TO of CGAs:

\begin{enumerate}
    \item Firstly, discrete element based representations are fundamentally limited in their ability to model continuous variations of the design.

    \item Secondly, the resolution of the design is constrained by the discrete mesh element size. The highest achievable design resolution without aliasing is bound by the Nyquist frequency associated with the mesh, imposing limitations on the attainable design complexity. 

    \item  Lastly, the calculation of spatial gradients, crucial for enforcing gradation constraint (\Cref{eq:opt_intro_maxGradationCons}), often relies on finite differences which can be inaccurate with a discretized representation.
\end{enumerate}

To address these limitations, we propose an implicit design representation based on band-limited NNs. Our approach builds upon the work \cite{chandrasekhar2021tounn, chandrasekhar2021multi, chandrasekhar2022LengthScale} which introduced neural network-based design representation for TO. The design is represented as a function of input coordinates $(\bm{x})$, with the network's weights $(\bm{w})$ acting as the design variables; $\vec{\rho} = \text{NN}(\bm{x}; \bm{w})$. This yields a mesh-independent representation, enabling design evaluation and optimization at arbitrary resolution. Further the representation allows for the querying of the design, its gradients, and higher-order derivatives throughout the domain.

Furthermore, as discussed in \Cref{sec:relatedWork}, we adopt a band limited network to represent our design. In particular, building up the  multiplicative filter network (MFNs) architecture \cite{fathony2020multiplicativeFilterNetworks} we make the following observations:

\begin{enumerate}
    \item The design is parameterized by the weights $(\bm{w})$ of the MFN.  This effectively decouples the design complexity from the underlying discretization. Furthermore, the design can be queried at any coordinate within the domain, making it suitable to represent CGAs. 
    \item MFNs allow for an explicit control over the frequency content of the design representation. This is crucial for ensuring that the design remains band-limited and avoids aliasing artifacts \cite{white2018FourierTO}.
    \item Further, MFNs allow us to tune the maximum represented frequency independent of the mesh resolution. Further, this also allows us to constrain the maximum gradation implicitly, removing the need for additional explicit constraints (\Cref{eq:opt_intro_maxGradationCons}).
\end{enumerate}

In addition to these improvements, our NN-based representation inherits the advantages of previous work \cite{chandrasekhar2021tounn}, including the ability to extract high-resolution designs, efficient optimization independent of mesh resolution, and end-to-end differentiability. Moreover, the network exhibits constrained behavior at unsupervised points, facilitating the use of a coarser mesh with a feature size exceeding that required by the gradation constraint \cite{sigmund1998numericalInstabilitiesTO}. This translates to accelerated finite element analysis and cost-effective optimization.

\subsection{Band Limited Coordinate Network}
\label{sec:method_baconNetwork}

Recall that our objective is to optimize CGA components that satisfy a maximum composition gradation constraint. Towards this, let us first examine Bernstein's Inequality:

\begin{theorem}[Bernstein's Inequality \cite{lapidoth2017foundationInDigitalComm}]
Let \(\rho(x) \in L^\infty(\mathbb{R})\) be a real-valued, bandlimited signal with bandwidth \(B > 0\); i.e., its Fourier transform vanishes outside the interval \([-\overline{B}, \overline{B}]\). If \(|\rho(x)| \leq M\) for all \(x \in \mathbb{R}\), then its derivative satisfies:

\begin{equation}
\left| \frac{d\rho}{dx} \right| \leq 2 \pi \overline{B}M \quad , \; \forall t \in \mathbb{R}
\label{eq:bernsteinInequality}
\end{equation}

\label{thm:bernsteinInequality}
\end{theorem}

The theorem captures the intuition that the rate at which a signal can change is proportional to its bandwidth. Considering our particular case, where  $|\rho^{(j)}(\bm{x})| \leq 1$ (\Cref{eq:opt_intro_boundCons}) and its derivative by $|d\rho^{(j)}/dx| \leq 1/L_{max}$ (\Cref{eq:opt_intro_maxGradationCons}), \Cref{thm:bernsteinInequality} implies that $\rho^{(j)}(\bm{x})$ must then be bandlimited with a bandwidth $\overline{B} = 1/2\pi L_{max} $. In other words, to ensure a maximum gradation over $L_{max}$ (\Cref{eq:opt_intro_maxGradationCons}), our design must be bandlimited, with a maximum frequency of $2\pi\overline{B}$.

\begin{figure}[H]
 	\begin{center}
		\includegraphics[scale=0.65,trim={0 0 0 0},clip]{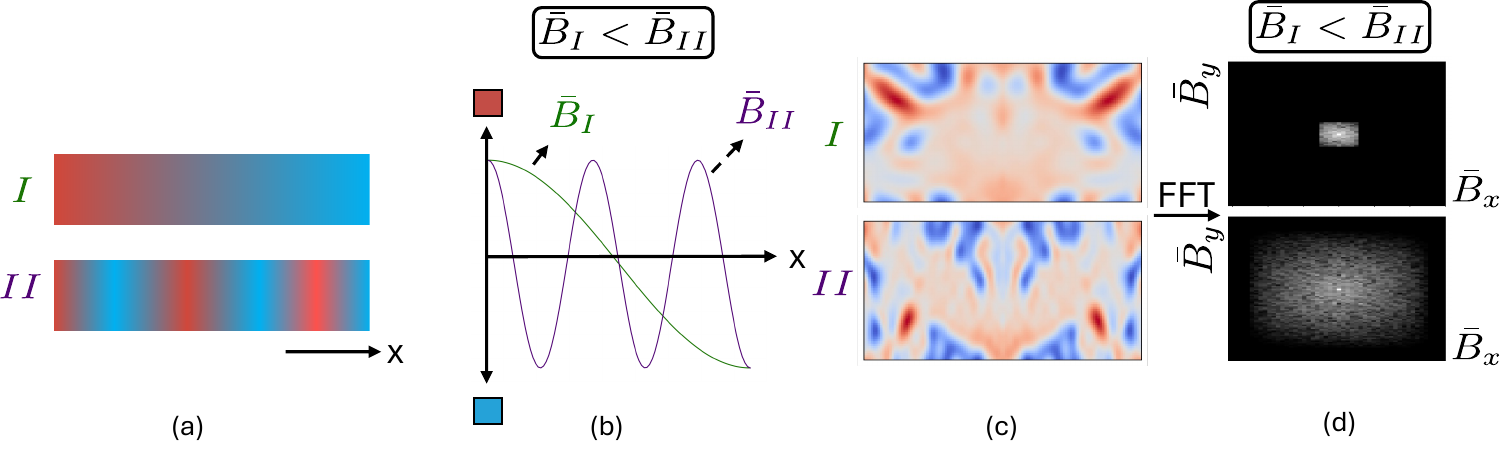}
 		\caption{(a) One-dimensional examples of gradual ($\Romannum{1}$) and sharp ($\Romannum{2}$) compositional variations. (b) The bandlimit for the gradual variation is lower than that for the sharp variation. (c) Two-dimensional optimized compositions exhibiting smooth ($\Romannum{1}$) and sharp ($\Romannum{2}$) variations. (d) The Fast Fourier Transform (FFT) of the corresponding composition fields. }
 		\label{fig:freq_description}
	\end{center}
 \end{figure}

\Cref{fig:freq_description} provides a visual intuition for this principle. We first consider a simple one-dimensional case with two materials. \Cref{fig:freq_description}(a) illustrates a gradual compositional variation ($\Romannum{1}$) and a much sharper, more abrupt variation ($\Romannum{2}$). It is intuitive that representing the sharp transition requires higher frequency components than the gradual one. This is confirmed in \Cref{fig:freq_description}(b), which shows that the band limit for the gradual variation is  lower than that for the sharp variation. This understanding extends directly to two-dimensional designs. \Cref{fig:freq_description}(c) displays two optimized compositions, one with smooth, gradual features ($\Romannum{1}$) and another with sharper, more detailed features ($\Romannum{2}$). We then compute the Fast Fourier Transform (FFT) of these corresponding composition fields, shown in \Cref{fig:freq_description}(d). The FFT of the smoother design reveals that its frequency content is concentrated in a small region around the center, indicating a low band limit. In contrast, the FFT of the sharper design shows a wider distribution of frequencies. Critically, in both examples, the frequency content is confined, and its magnitude becomes negligible outside of the imposed band limit. This demonstrates that the optimized designs are indeed band-limited.

\begin{figure}[h]
 	\begin{center}
		\includegraphics[scale=0.5,trim={0 0 0 0},clip]{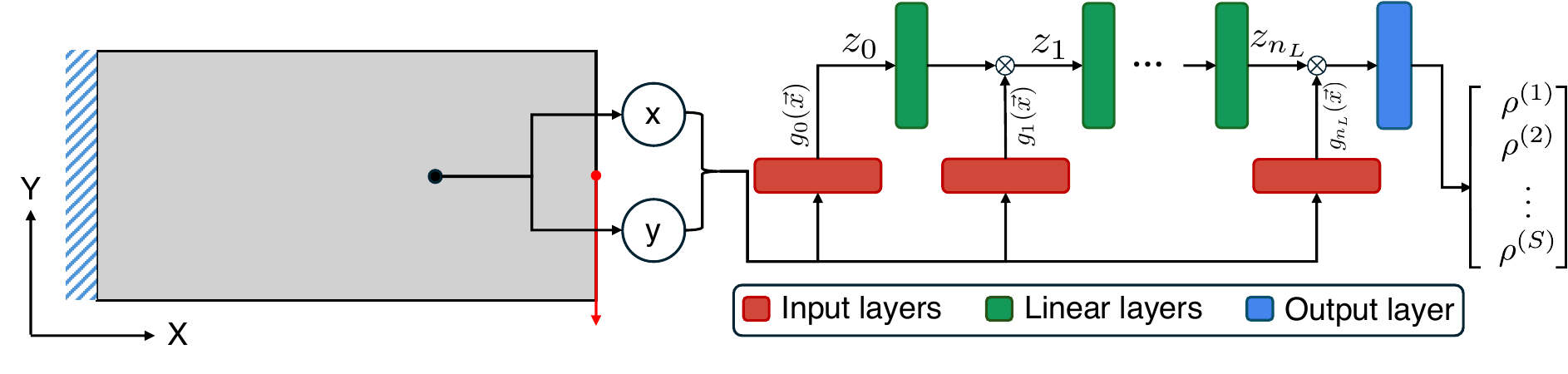}
 		\caption{The neural network maps input coordinates to component fractions.}
 		\label{fig:nn_architecture}
	\end{center}
 \end{figure}

With this understanding, we aim to design a network that is implicitly band-limited with a maximum frequency of $\overline{B}$. Our proposed network is inspired by the architecture presented in \cite{fathony2020multiplicativeFilterNetworks, lindell2022bacon, shekarforoush2022rRMFN}. We briefly summarize the construction and key features of this network, referring readers to \cite{lindell2022bacon} for a comprehensive discussion, including architecture, spectral analysis, and network initialization. Notably, while conventional MLPs \cite{sitzmann2020SIREN} consist of one input layer connected to a series of hidden layers, the MFN comprises multiple input and hidden layers (\Cref{fig:nn_architecture}). $\textit{The gradation control is achieved by tuning the frequencies of the input layers}$. Specifically, for an MFN consisting of  $n_L$ layers, each containing $d_h$ neurons, we have:

\begin{enumerate}
    \item $\textbf{Input layers :}$ The spatial coordinates $\bm{x} \in \mathbb{R}^{d_{in}}$ are transformed by a series of sine-activated input layers, each of the form $h_i : \mathbb{R}^{d_{in}} \rightarrow \mathbb{R}^{d_h}$ , where $h_i(x) = sin(\omega_i \bm{x}  + \phi_i)$, for $ i=0,\ldots,n_L - 1$. The initialization of the parameters $\omega_i \in \mathbb{R}^{d_h \times d_{in}}$ and $\phi_i \in \mathbb{R}^{d_h} $ is crucial to this work, as these values control the bandwidth of the output signal. Specifically, with  $\omega_i \in (-B_i, B_i)$ and $\phi_i \in (-\pi, \pi)$ the total bandwidth is the sum of the individual bandwidths. To satisfy the gradation constraint, we require $\sum\limits_{i=0}^{n_L - 1} B_i = \overline{B}$.

    \item $\textbf{Hidden layers :}$ The outputs of the input layers are then multiplied element-wise with the output of the preceding input layer after passing through a linear transformation. These Hadamard products of sinusoids generate summed frequencies, yielding an exponential number ($\mathcal{O}(d_h^{n_L})$ ) of sine bases while maintaining a computationally efficient polynomial number of parameters ($\mathcal{O}(n_L d_h^2)$). In particular with $\mathbf{z}_0 = h_0(\mathbf{x}) $, we have:

        \begin{equation}
        \mathbf{z}_i = h_i(\mathbf{x}) \odot (\mathbf{W}_i \mathbf{z}_{i-1} + \mathbf{b}_i), \quad 0 < i < N_L
        \label{eq:hiddenLayerMFN}
        \end{equation}
    
    where, $\mathbf{W}_i \in \mathbb{R}^{d_h \times d_h}$ and $\mathbf{b}_i \in \mathbb{R}^{d_h}$ are the design parameters associated with the hidden layers.
    
    \item $\textbf{Output layer :}$ Finally, the output of the last hidden layer is passed through a linearly activated output layer comprising $S-1$ neurons to yield the component fractions at the input coordinates.
        \begin{equation}
            \bm{\rho} = \mathbf{W}_\text{out} \mathbf{z}_{n_L - 1} + \mathbf{b}_\text{out}
            \label{eq:outputLayerMFN}
        \end{equation}
    where $\mathbf{W}_\text{out} \in \mathbb{R}^{d_h \times (S)}$ and $\mathbf{b}_\text{out} \in \mathbb{R}^{S-1}$ are the design parameters associated with the output layer. Note that the output of the NN is unbounded, i.e., $\bm{\rho} \in \mathbb{R}$. Ensuring that the obtained $\bm{\rho}$ are physically valid (i.e., satisfy \Cref{eq:opt_intro_boundCons}) is addressed in Section \Cref{sec:method_boundCons}.
\end{enumerate}

\subsection{Material Model}
\label{sec:method_materialModel}

 \begin{figure}[h]
 	\begin{center}
		\includegraphics[scale=0.4,trim={0 0 0 0},clip]{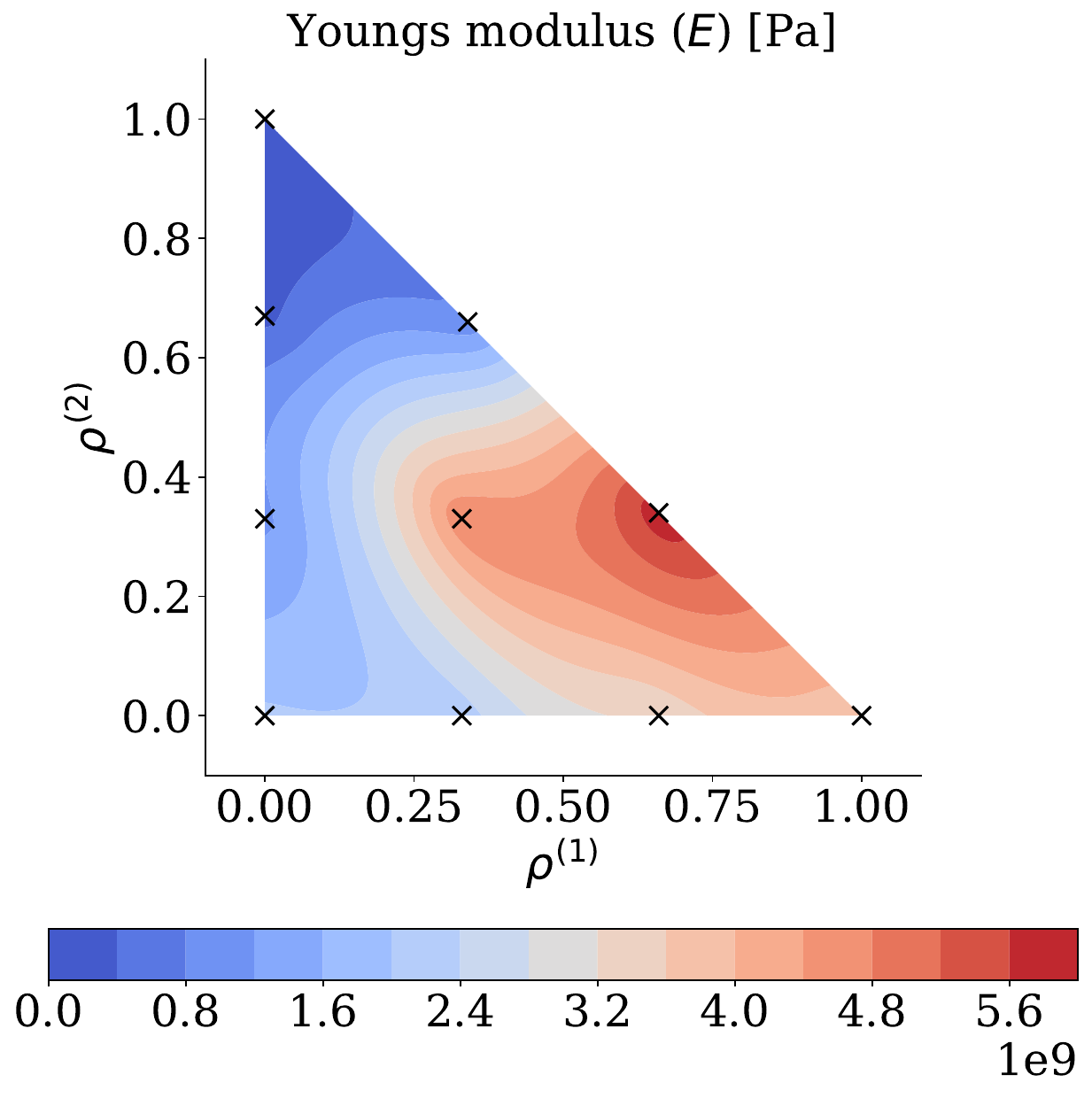}
 		\caption{The material properties are determined at a set of component compositions $(\times)$ and interpolated throughout the composition space using radial basis functions. Illustration of a sample ternary material system.}
        \label{fig:rbf_interpolation_example}
	\end{center}
 \end{figure}

Having obtained the spatial distribution of the composition, we now obtain the distribution of the material properties. In particular, material properties relevant to our design objective, such as Youngs modulus and mass density, are composition dependent. We assume that these relevant properties have been determined either through high throughput experimentation \cite{vecchio_high-throughput_2021, Niklas2023-highthroughput, feng_high-throughput_2021} or through numerical simulation (e.g., CALPHAD \cite{lukas2007CALPHAD}) for a set of compositions. For instance, let us suppose $\{E^*_1, E^*_2, \ldots, E^*_M\}$ are the Youngs modulus at $M$ corresponding compositions $\{ \vec{\rho}_1^{*}, \vec{\rho}_2^{*}, \ldots, \vec{\rho}_M^{*}\}$. Then, the modulus at arbitrary intermediate composition  $\bar{E}(\vec{\rho})$ are interpolated using a radial basis function (RBF) \cite{press2007numericalRecipes} as:

\begin{equation}
    \bar{E}(\vec{\rho}) = \sum\limits_{j=1}^{M} \tilde{E}_j \Phi(|| \vec{\rho} - \vec{\rho}_j^* ||)
    \label{eq:RBF_interp_youngs_modulus}
\end{equation}

where $\tilde{E}_j$ are the associated RBF parameters derived from $E^*$ and $\vec{\rho}^{*}$. Further, $\Phi(\cdot)$ is the Gaussian RBF kernel \cite{press2007numericalRecipes}. For illustration, see \Cref{fig:rbf_interpolation_example}. Here, the material properties are determined at compositions $(\times)$ for an idealized ternary material system and are interpolated at all other compositions. Similar expressions follow for other relevant properties such as the mass density $(\lambda)$, thermal conductivity $(\kappa)$, and thermal expansion coefficient $(\alpha)$.

\subsection{Finite Element Analysis}
\label{sec:method_fea}

This section details the process for determining the system's response to applied loads. We focus on solving the governing partial differential equations (PDEs) for structural mechanics. We also consider a weakly coupled thermo-elastic system where temperature changes induce thermal strains. The structural response is governed by the following PDE:

\begin{subequations}
\begin{align}
\bm{R}_u &\coloneqq \sigma_{ij,j} + b_j = 0 \quad \text{in } \Omega \\
\sigma_{ij} &= E_{ijkl}(\epsilon_{kl} - \alpha(T - T_0)\delta_{kl}) \quad \text{in } \Omega \\
\epsilon_{ij} &= \frac{1}{2}(u_{i,j} + u_{j,i}) \quad \text{in } \Omega \\
u_i &= \bar{u}_i \quad \text{on } \Gamma_u \\
\sigma_{ij}n_j &= \bar{t}_j \quad \text{on } \Gamma_t
\end{align}
\label{eq:struct_PDE}
\end{subequations}

Here, $\sigma_{ij}$ is the stress tensor, $b_j$ is the body force vector, and $E_{ijkl}$ is the elasticity tensor. $\epsilon_{kl}$ is the strain tensor, and $u_i$ is the displacement on the Dirichlet boundary $\Gamma_u$. For thermo-elastic problems, $\alpha$ is the coefficient of thermal expansion, $T$ is the current temperature, $T_0$ is a reference temperature, and $\bar{t}_j$ is the traction on the Neumann boundary $\Gamma_t$.

The temperature field $T$ is found by solving the thermal PDE:

\begin{subequations}
\begin{align}
\bm{R}_T &\coloneqq -q_{j,j} + s = 0 \quad \text{in } \Omega  \\
q_j &= -\kappa T_{,j} \quad \text{in } \Omega  \\
T &= \bar{T} \quad \text{on } \Gamma_T \\
q_j n_j &= \bar{q} \quad \text{on } \Gamma_q
\end{align}
\label{eq:thermal_PDE}
\end{subequations}

Here, $q_j$ is the heat flux vector, $s$ is a body heat source, and $\kappa$ is the thermal conductivity. $\bar{T}$ is the prescribed temperature on $\Gamma_T$, and $\bar{q}$ is the prescribed heat flux on $\Gamma_q$.

We solve the PDEs using a multigrid finite element (FE) solver \cite{Bell2022PyAMG}. The solution process is as follows. First, we evaluate material properties at the center of each element. These properties are Young's modulus $E(\vec{\rho}_e)$, thermal conductivity $\kappa(\vec{\rho}_e)$, and the thermal expansion coefficient $\alpha(\vec{\rho}_e)$.

For a thermo-elastic analysis, we first solve the thermal problem in \Cref{eq:thermal_PDE}. We use $\kappa(\vec{\rho}_e)$ to assemble the thermal stiffness matrix, $\bm{K}_T$. The resulting temperature field, $\bm{T}$, and the coefficient $\alpha(\vec{\rho}_e)$ are used to compute the thermal forces. Next, we solve the structural problem in \Cref{eq:struct_PDE}. We use $E(\vec{\rho}_e)$ to assemble the mechanical stiffness matrix, $\bm{K}_m$. The total force vector, $\bm{f}$, includes mechanical forces, $\bm{f}_m$, and any thermal forces. Solving the system yields the displacement field, $\bm{u}$. Finally, we compute the objective (for instance structural compliance, $J$), from the obtained displacement field $\bm{u}$. For a more detailed formulation, we refer the reader to \cite{thurier2019twoMaterialThermoElasticTO,shishir2024multiMaterialThermoElasticTO}.

\subsection{Mass constraint}
\label{sec:method_massCons}

A primary objective of our formulation is to achieve optimally performant lightweight structures. To this end, we impose a constraint on the mass of the optimized design. Following \Cref{sec:method_materialModel}, we derive the mass densities at the element centers $(\lambda(\vec{\rho}_e))$. Subsequently, with $A_e$ being the area of the elements and $m^*$ being the maximum allowed mass of the structure, we can express the mass constraint $(g_m)$ as:

\begin{equation}
    g_m(\bm{\vec{\rho}}) \coloneqq \sum\limits_{e=1}^{N_e} \lambda(\vec{\rho}_e) A_e \leq m^*
    \label{eq:mass_cons}
\end{equation}

\subsection{Bound and Partition of Unity Constraint}
\label{sec:method_boundCons}

Recall that to ensure the physical validity of component fractions, it is necessary to bound $\rho^{(j)} \in [0,1] \; , \; \forall j$ (\Cref{eq:opt_intro_boundCons}) and enforce partition of unity (\Cref{eq:opt_intro_partitionUnityCons}). Recall that the output of the NN unbounded \Cref{sec:method_baconNetwork}. Although activation functions such as sigmoid function can be introduced at the output layer to bound the output to $[0,1]$ \cite{chandrasekhar2021tounn}, it introduces unregulated and higher frequencies, compromising the band-limited nature of the network \cite{tancik2020fourier}. Instead, we introduce explicit constraints to enforce \Cref{eq:opt_intro_boundCons,eq:opt_intro_partitionUnityCons}.

Observe that we enforce $0 \leq \rho_e^{(j)} \leq 1 $ for each observed point ($ \forall \bm{x}_e$), and component ($\forall j$). This results in a large number of constraints. We thus simplify by aggregating them into two constraints. Specifically, the aggregated upper bound constraint can expressed as:

\begin{equation}
    g_u(\vec{\bm{\rho}}) \coloneqq \underset{e, j}{\text{max}}(\rho_e^{(j)}) <= 1
    \label{eq:upper_bound_cons}
\end{equation} 
and the lower bound as:

\begin{equation}
    g_l(\vec{\bm{\rho}}) \coloneqq \underset{e, j}{\text{min}}(\rho_e^{(j)}) >= 0
    \label{eq:lower_bound_cons}
\end{equation}

To maintain differentiability, we employ the LogSumExp (LSE) approximation for the maximum and minimum functions \cite{zhang2021dive}. For instance, we reexpress \Cref{eq:upper_bound_cons}
in \Cref{eq:upper_bound_cons_lse}. Similar expression follows for the lower bound constraint \Cref{eq:lower_bound_cons}.

\begin{equation}
    g_u(\vec{\bm{\rho}}) \coloneqq  \frac{1}{t} \log \left( \sum_{j=1}^{S}\sum_{e=1}^{N_e} \exp\left(t \rho_e^{(j)}\right) \right)  - 1\leq 0
    \label{eq:upper_bound_cons_lse}
\end{equation}
where, $t$ is the LSE scaling factor (=10 in our experiments). Furthermore, we express the imposed partition of unity constraint as:

\begin{equation}
    g_p(\bm{\vec{\rho}}) \coloneqq \frac{1}{N_e} \sum\limits_{j=1}^S \rho^{(j)}(\bm{x}_e) - 1 = 0
    \label{eq:partition_unity_cons}
\end{equation}
The bound constraint along with the partition of unity of constraint ensures that the obtained material compositions are physically valid.

\subsection{Optimization problem}
\label{sec:method_optimization}

With the objective (\Cref{sec:method_fea}), constraints (\Cref{sec:method_massCons,sec:method_boundCons}), and design variables (i.e., the weights of the network; \Cref{sec:method_baconNetwork}) defined, we can now reformulate the optimization problem in \Cref{eq:optimizationIntroEquations} as \Cref{eq:optimizationEquations}. Note that the governing residual equations  are implicitly satisfied when solving for the temperature and displacement fields through the FE formulation (\Cref{sec:method_fea}). Additionally, the mass constraint (\Cref{eq:opt_intro_massCons}) is computed as detailed in \Cref{sec:method_massCons}. Furthermore, the bound and partition of unity constraints in \Cref{eq:opt_intro_boundCons,eq:opt_intro_partitionUnityCons} are computed as detailed in \Cref{sec:method_boundCons}. Finally, the maximum gradation constraint (\Cref{eq:opt_intro_maxGradationCons}) is met by the choice of network frequencies (\Cref{sec:method_baconNetwork}).

\begin{subequations}
	\label{eq:optimizationEquations}
	\begin{align}
		& \underset{\bm{w}} {\text{minimize}}  & J(\bm{w})
            \label{eq:opt_objective} \\
		& \text{such that} & \bm{R}_T= 0
            \label{eq:opt_govnEqThermal} \\
            & & \bm{R}_u = 0  
            \label{eq:opt_govnEqStruct} \\
            & &  g_m(\bm{w}) \leq 0
            \label{eq:opt_massCons} \\
            & &  g_u(\bm{w}) \leq 0
            \label{eq:opt_upperBoundCons} \\
            & &  g_l(\bm{w}) \leq 0
            \label{eq:opt_lowerBoundCons} \\
            & &  g_p(\bm{w}) = 0
            \label{eq:opt_partitionCons}
	\end{align}
\end{subequations}

The constrained minimization problem \Cref{eq:optimizationEquations} is then transformed into an  unconstrained  loss  function  minimization,  using  the log-barrier scheme \cite{kervadec2022LogBarrier, nocedal1999Optimization}. Specifically, the loss function is defined as:

\begin{equation}
    \mathcal{L}(\mathbf{w}) = \frac{J}{J^0} + \psi(g_m) + \psi(g_u) + \psi(g_l) + \psi(g_{p^+}) + \psi(g_{p^-})
    \label{eq:net_loss}
\end{equation}

Where,

\begin{equation}
\psi_{\tau}(g) =
        \begin{cases}
        -\frac{1}{\tau} \ln(-g), & g \leq -\frac{1}{\tau^2} \\
        \tau g - \frac{1}{\tau}\ln(\frac{1}{\tau^2}) + \frac{1}{\tau}, & \text{otherwise}
        \end{cases}
\label{eq:loss_log_barrier}
\end{equation}

 and $J^0$is the initial objective. The constraint penalty parameter $\tau$ is updated at each iteration $'k'$ as $\tau = \tau_0 \mu^k$ (where, $\tau_0 = 3$ and $\mu = 1.04$), making the enforcement of the constraint stricter as the optimization progresses. The gradient-based Adam optimizer \cite{kingma2014adam} is used to minimize \Cref{eq:net_loss}. Note that the log-barrier treats the equality constraint $g_p$ as two constraints $g_{p^+} \coloneqq g_p \leq 0 $ and $g_{p^-} \coloneqq -g_p \leq 0 $.

Finally, to update the design parameters $\bm{w}$, we require the gradients of the loss with respect to these parameters. Traditionally, this is conducted manually, which can be labor-intensive and error-prone. However, leveraging the automatic differentiation (AD) capabilities \cite{chandrasekhar2021auto, xue2023jaxFem} of the JAX framework, we automatically derive the sensitivity of the loss to the network weights; $\frac{\partial \mathcal{L}}{\partial \bm{w}}$. In practice, this means we only need to define the forward expressions, and JAX's $\textit{autograd}$ library \cite{jax2018github} computes all necessary derivatives with machine precision. Finally, utilizing a NN-based design representation with ML-based optimizers, we harness the computational efficiency of tailored software and hardware, accelerating the design process.

\subsection{Algorithm}
\label{sec:method-algo}

 \begin{figure}[h]
 	\begin{center}
		\includegraphics[scale=0.5,trim={0 0 0 0},clip]{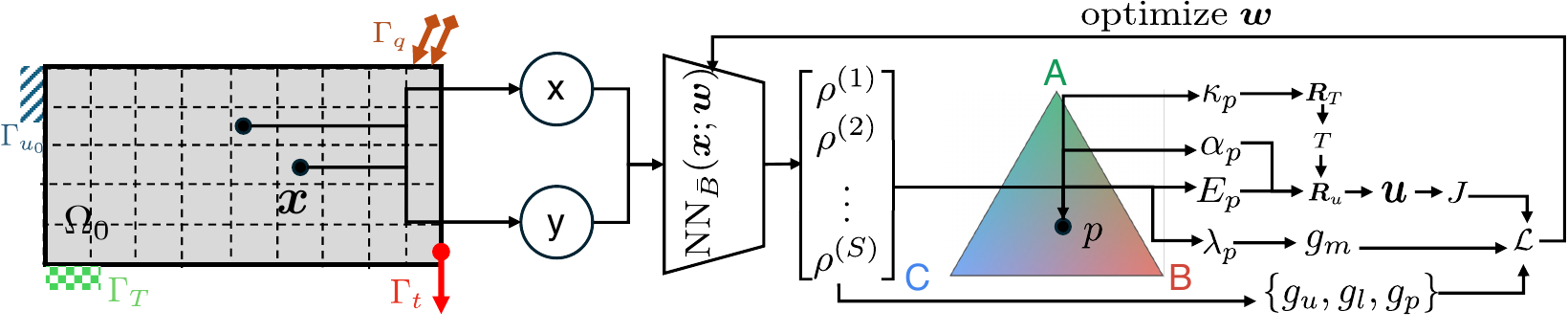}
 		\caption{Optimization loop of the proposed framework.}
 		\label{fig:flowchart_tobaco}
	\end{center}
 \end{figure}

The algorithm for the proposed framework is illustrated in \Cref{fig:flowchart_tobaco}. We begin by assuming that a thermo-elastic TO problem, a maximum allowed composition gradation and design mass, a material system with properties at a set of component fractions, and a NN configuration have been prescribed.

\begin{enumerate}
    \item \textbf{Domain discretization :} The prescribed domain is discretized, with the mesh utilized for the thermal and structural FEA. The center coordinates of the mesh elements act as inputs to our NN.

    \item \textbf{Neural network initialization :} The NN is initialized with the bandwidth as determined from the allowed maximum gradation. The network weights are initialized such that the output for any coordinate within the domain results in all output component fractions being equal to $1/S$. 

    \item \textbf{Material property Initialization :} The RBFs, which are used to interpolate the properties throughout the material space, are initialized based on the available material property data.
\end{enumerate}

After initialization, the optimal material distribution is computed by iterating through the main optimization loop till convergence (or till the number of iterations exceed a set maximum limit):

\begin{enumerate}

    \item \textbf{Component fraction Computation :} The NN predicts the proportion of $S$ components $\{\rho^{(1)}, \ldots, \rho^{(S)}\}$ at the center of each element $\bm{x}_e$. The network is parameterized by the weights $\bm{w}$ which constitute our design variables for the optimization.

    \item \textbf{Material property estimation :}  The RBF interpolators are used to determine the relevant material properties at each element center from the computed component fractions. For instance \Cref{fig:flowchart_tobaco} illustrates a Fe-Cr-Ni material system from which the thermal conductivity $\kappa$, thermal expansion coefficient $\alpha$, Youngs modulus $E$ and mass density $\lambda$ are obtained at the component fraction $p$.

    \item \textbf{Thermal FEA :} The thermal PDE ($\bm{R}_T$) is solved using FEA to compute the temperature distribution $T$ within the design under the applied thermal loads. The elastic strain arising from thermal expansion are determined based on the obtained temperature field and material properties.

    \item \textbf{Structural FEA :} The structural PDE ($\bm{R}_u$) is then solved using FEA to compute the displacement field $\bm{u}$ under both the thermo-elastic forces and any applied structural loads.

    \item \textbf{Objective and constraint evaluation :} The objective $(J)$, mass $(g_m)$, bound $(g_l, g_u)$ and partition constraints $(g_p)$ are computed. The net loss $(\mathcal{L})$, combining the objective and constraint violations, is then determined.

    \item \textbf{Sensitivity analysis :}  Automatic differentiation is employed to compute the sensitivity of the net loss to the weights of the NN (design variables).

    \item \textbf{Weight update :}  With the sensitivities computed, the NN weights are updated using the Adam optimizer. 

    \item \textbf{Penalty parameter updates :} Following parameter continuation schemes \cite{rojas2015ContinuationStrategyTO}, the penalty parameters are adjusted to gradually increase the enforcement of the constraints as the optimization progresses.
    
\end{enumerate}

\section{Numerical Experiments}
\label{sec:expts}

In this section, we conduct several experiments to illustrate the proposed framework. All experiments are conducted on using the JAX library \cite{jax2018github} in Python; in particular the JAX-FEM library \cite{xue2023jaxFem}. The default parameters used in the experiments are as follows:

\begin{enumerate}
    \item \textbf{Neural network :} A network with 3 hidden layers; with 100 neurons in each layer is used.
    \item \textbf{Mesh :} A structured bilinear quad mesh with $120 \times 60$ elements is used. A domain of size $2 \times 1$ is used.
    \item \textbf{Optimizer :} An ADAM optimizer with a learning rate of $10^{-2}$ is used. To improve stability, a gradient clip with a norm threshold of 1.
    \item \textbf{Convergence :} The optimization is terminated either till a maximum of 500 iteration or when $\Delta\mathcal{L} \leq 10^{-3}$.
    
\end{enumerate}

\subsection{Validation of Hypothesis}
\label{sec:expts_validation}

\begin{figure}[h]
 	\begin{center}
		\includegraphics[scale=0.26,trim={0 0 0 0},clip]{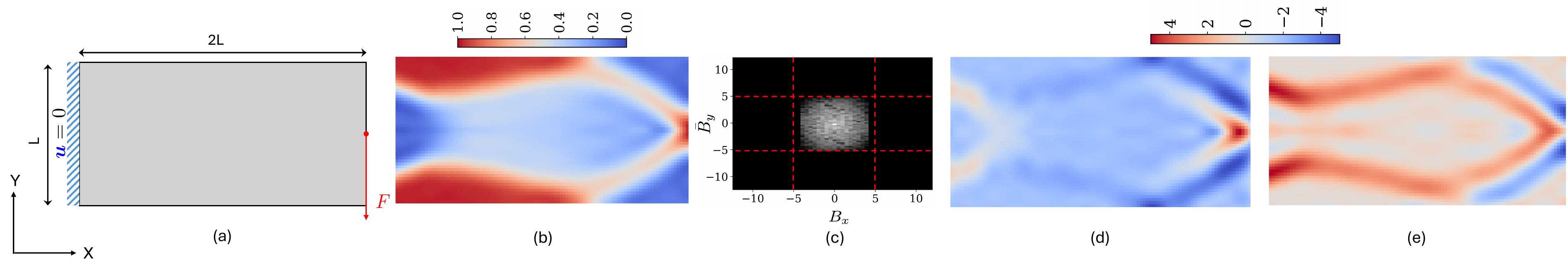}
 		\caption{(a) The domain and boundary conditions (b) Optimized material distribution (c) FFT of the material distribution  with markers at $\bar{B} = 5$  (d) Composition gradient along X (e) Composition gradient along Y.}
 		\label{fig:validation_phase_fraction}
	\end{center}
 \end{figure}

In the first experiment, we validate the central hypothesis of the proposed method; where we can utilize a band limited neural network to represent a design and consequently control the gradation of the optimized component by restricting the bandwidth. For simplicity, we begin by considering a  structural compliance minimization problem in a two-dimensional domain under the loading conditions depicted in \Cref{fig:validation_phase_fraction}(a).

Furthermore, for simplicity, our analysis employs a single-component fraction, where the Young's modulus and mass density vary linearly with this component fraction; i.e., $E(\rho) = E_0 \rho$ and $\lambda(\rho) = \lambda_0 \rho$ respectively; with $E_0 = 1$ and $\lambda_0 = 1$. Additionally, we impose a gradation constraint with an upper limit $\bar{B} = 5$. The design optimization is conducted subject to a maximum allowable mass of $1$.

The resulting optimized material distribution is illustrated in \Cref{fig:validation_phase_fraction}(b), achieving a final compliance value of 63.6. Moreover, we conduct a FFT analysis of the optimized design, with the frequency spectrum displayed in \Cref{fig:validation_phase_fraction}(c). The analysis indicates that the resulting design is indeed band-limited, confirming the ability of the network to produce band limited designs. To evaluate the validity of the imposed gradation constraint, we compute the gradations by backpropagating through the network. The computed gradations, presented in \Cref{fig:validation_phase_fraction}(d) and (e), clearly adhere to the imposed constraint, thereby confirming our central hypothesis.

\begin{figure}[h]
 	\begin{center}
        \includegraphics[scale=0.5,trim={0 0 0 0},clip]{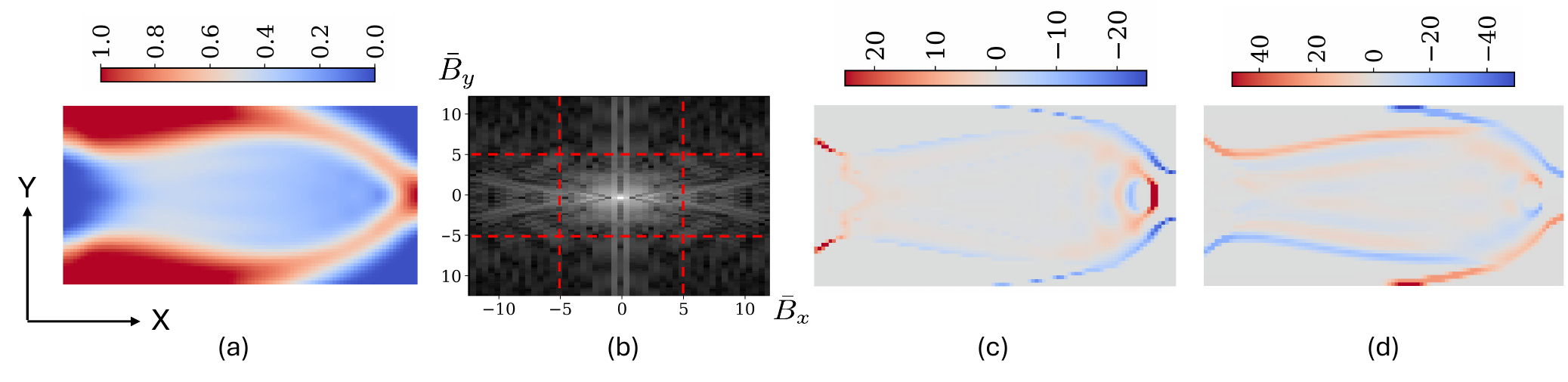}
 		\caption{The (a) optimized material distribution (b) the FFT with markers at $\bar{B} = 5$  (c) gradation along X (d) and Y axis using standard element based formulation.}
        \label{fig:mid_cant_density_TO}
	\end{center}
 \end{figure}

For comparison, we also illustrate the material distribution and its FFT following standard element based density based formulation in \Cref{fig:mid_cant_density_TO}. Observe that a lower compliance of 54.8 is achieved owing to the unrestricted freedom in its gradation. Note that the unlike standard TO we do not impose any penalization on the material stiffness, allowing for intermediate material compositions.

\begin{figure}[h]
 	\begin{center}
        \includegraphics[scale=0.3,trim={0 0 0 0},clip]{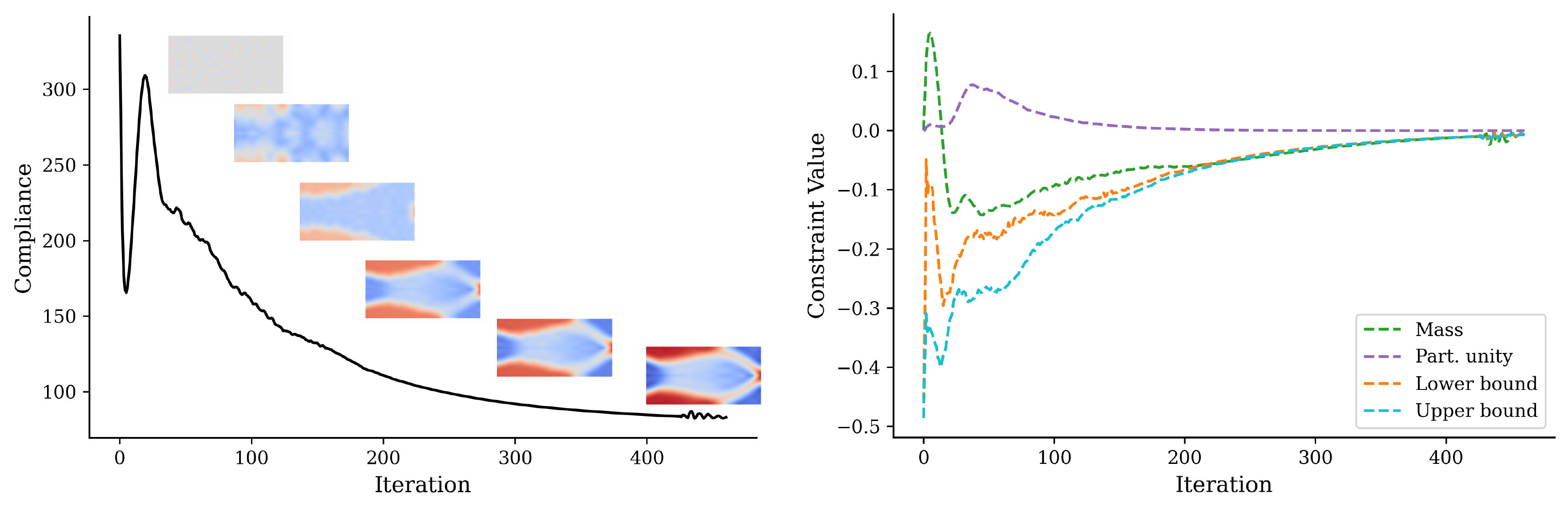}
 		\caption{Convergence of the objective (a) and constraints (b). The insets show the design at the $0^{th}$, $10^{th}$, $50^{th}$, $100^{th}$ $200^{th}$ and final iteration for the proposed algorithm.}
        \label{fig:convergence_tobaco}
	\end{center}
 \end{figure}

Finally, \Cref{fig:convergence_tobaco} illustrates the convergence of the objective function, and constraints for the same problem. The figure also shows the evolution of the  material distribution during optimization. The optimization took 461 iterations to converge; with a wall-clock time of 211 seconds. The stable convergence in the plot highlights the effectiveness and robustness of our proposed log-barrier formulation. Similar convergence characteristics were observed in the other experiments.

\subsection{Variation in Gradation limit}
\label{sec:expts_gradation}

 \begin{figure}[h]
 	\begin{center}
		\includegraphics[scale=0.2,trim={0 0 0 0},clip]{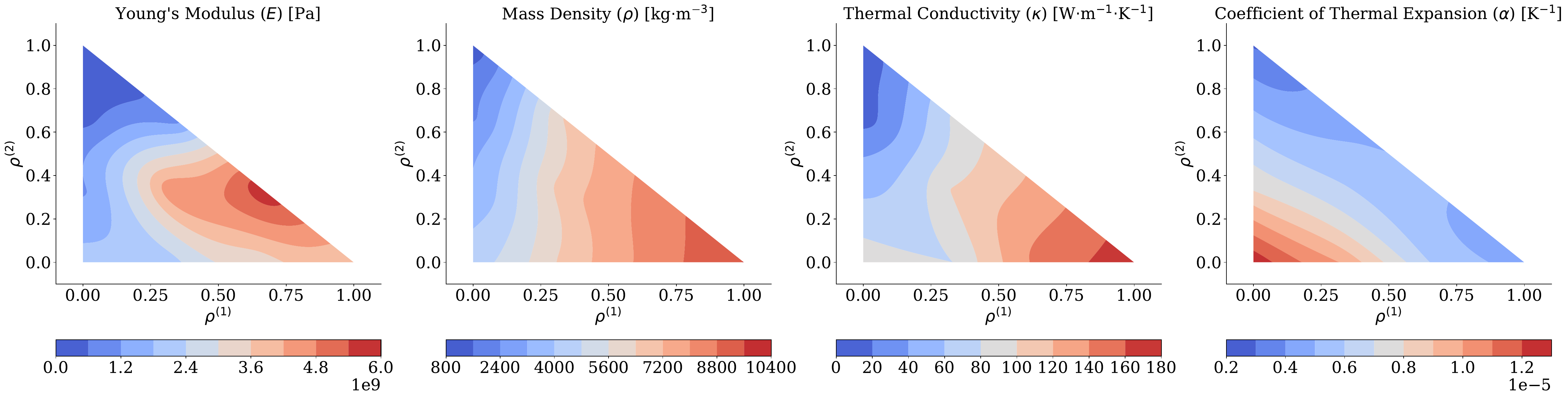}
 		\caption{Properties of a Ternary material system.}
 		\label{fig:material_properties}
	\end{center}
 \end{figure}

\begin{figure}
 	\begin{center}
		\includegraphics[scale=0.4,trim={0 0 0 0},clip]{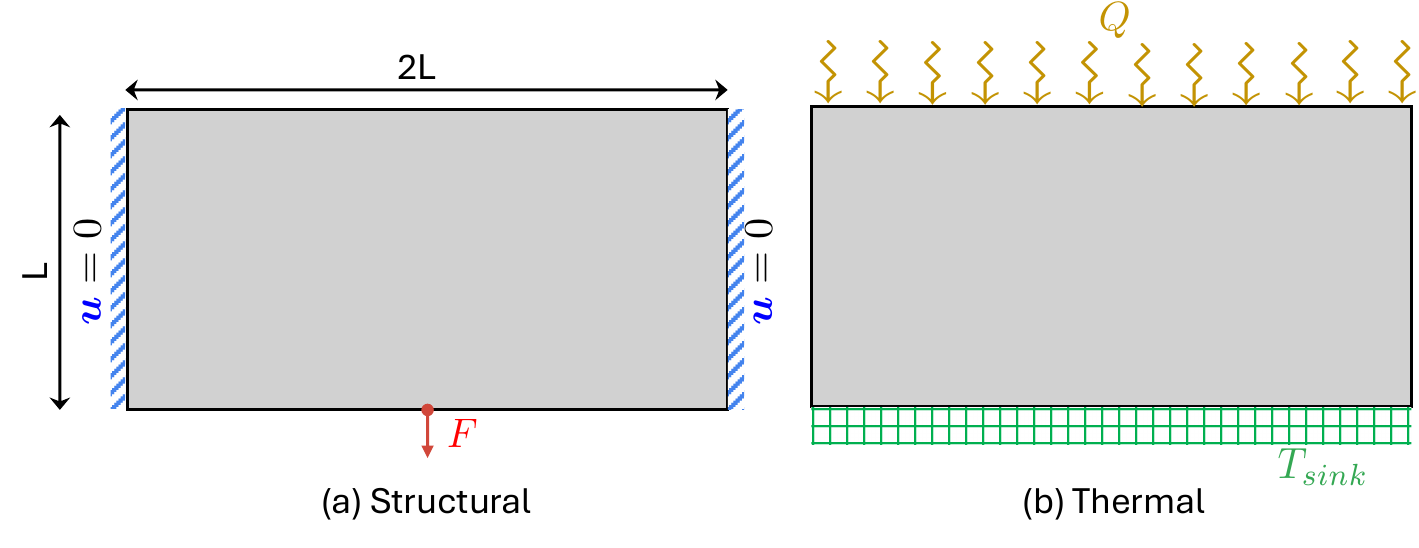}
 		\caption{Thermo-elastic boundary conditions.}
 		\label{fig:thermo_elastic_bc}
	\end{center}
\end{figure}

In this experiment, we study the effect of the gradation limit on optimization. We consider a three-component material system whose properties as a function of composition, is illustrated in \Cref{fig:material_properties}. The optimization addresses the classic thermo-elastic TO  problem described in \cite{rodrigues1995ThermoElasticProblem}, aiming to minimize mechanical compliance under thermally induced stresses. The domain and  boundary conditions are illustrated in \Cref{fig:thermo_elastic_bc}.  With a maximum allowed mass of 1kg and all other parameters held constant, we vary the gradation limit.

\begin{figure}[h]
 	\begin{center}
		\includegraphics[scale=0.5,trim={0 0 0 0},clip]{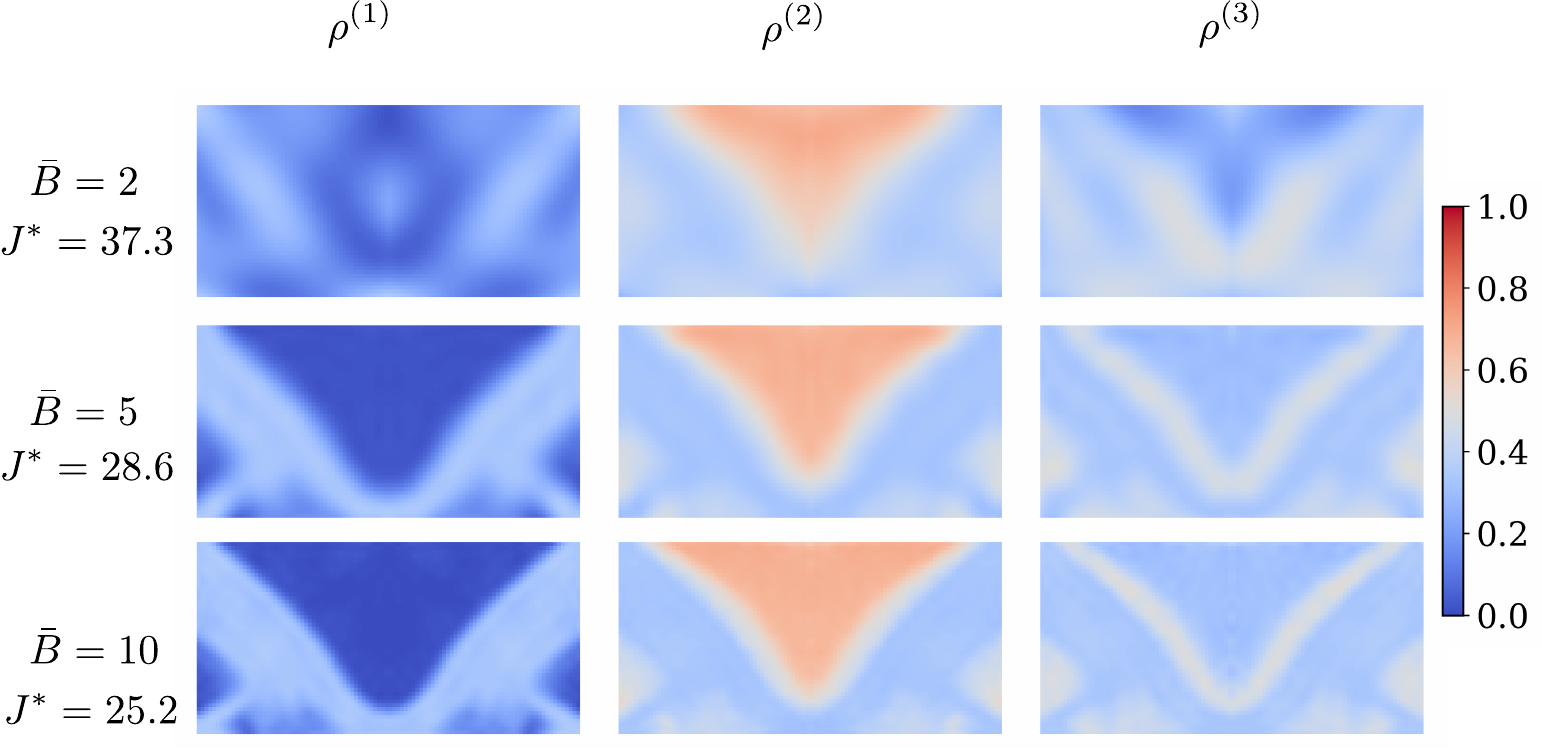}
 		\caption{Impact of network bandwidth (gradation limit) on the optimized design.}
 		\label{fig:gradation_variation}
	\end{center}
 \end{figure}

\Cref{fig:gradation_variation} illustrates the material composition of the three components at band limits of 2, 5, and 10. The compliances are 37.3, 28.6, and 25.2 respectively; indicating that compliance decreases as the bandwidth increases, corresponding to an increased design space. Furthermore, the designs become increasingly graded as the gradation limit increases, validating our premise. For compactness, the gradation and FFT plots are not illustrated, but this observation are consistent with with our previous experiment.

\begin{figure}[h]
 	\begin{center}
		\includegraphics[scale=0.5,trim={0 0 0 0},clip]{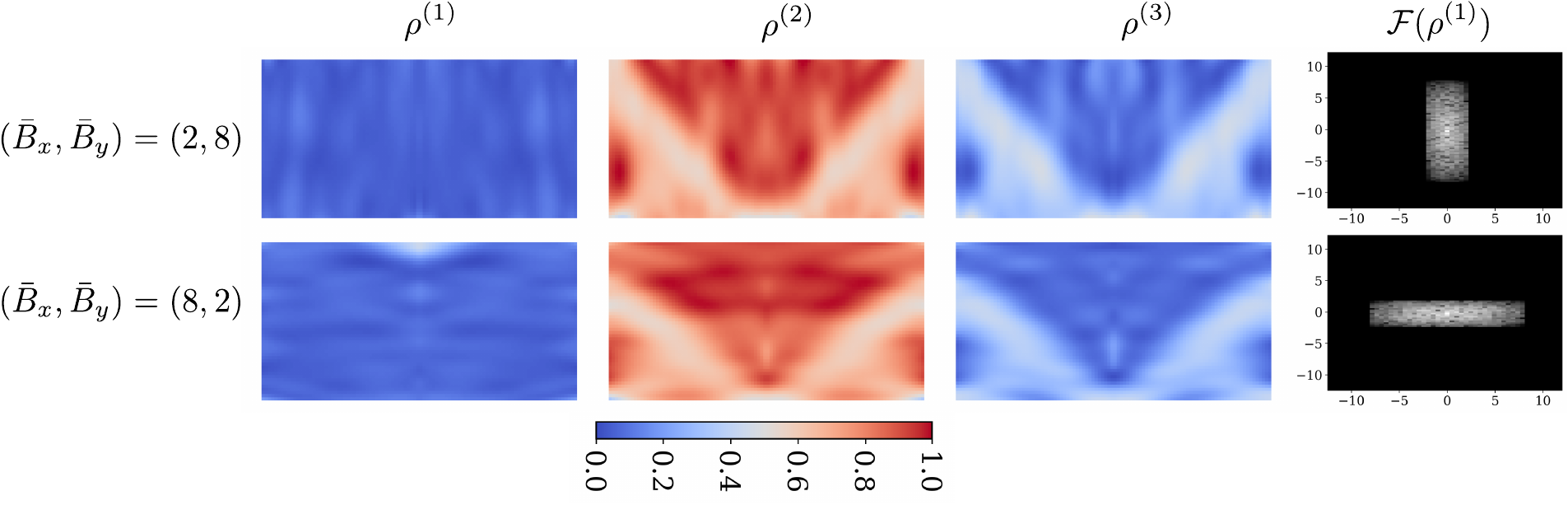}
 		\caption{Different gradation limit along the axes.}
 		\label{fig:directional_variation}
	\end{center}
 \end{figure}

Additionally, we address how certain AM processes such as material jetting allow compositional variation both within the printing plane and along the build direction, whereas processes like SLS restrict variation to the build direction alone. We can accommodate these differences by incorporating different band limits along different axes. To simulate such directional constraints, we limit the bandlimit of $(\bar{B}_x, \bar{B}_y) = (2, 8)$  and $(\bar{B}_x, \bar{B}_y) = (8, 2)$ in \Cref{fig:directional_variation}. The compliances are 31.3 and 35.7 respectively; indicating that allowing for variations along Y axis has a broader impact on the performance.  We also illustrate the FFT for $\rho^{(1)}$ in these cases to illustrate the non-symmetric bandlimit along the X and Y axes.

\subsection{Design of a Turbine Blade}
\label{sec:expts_turbineBlade}

\begin{figure}[h]
 	\begin{center}
		\includegraphics[scale=0.3,trim={0 0 0 0},clip]{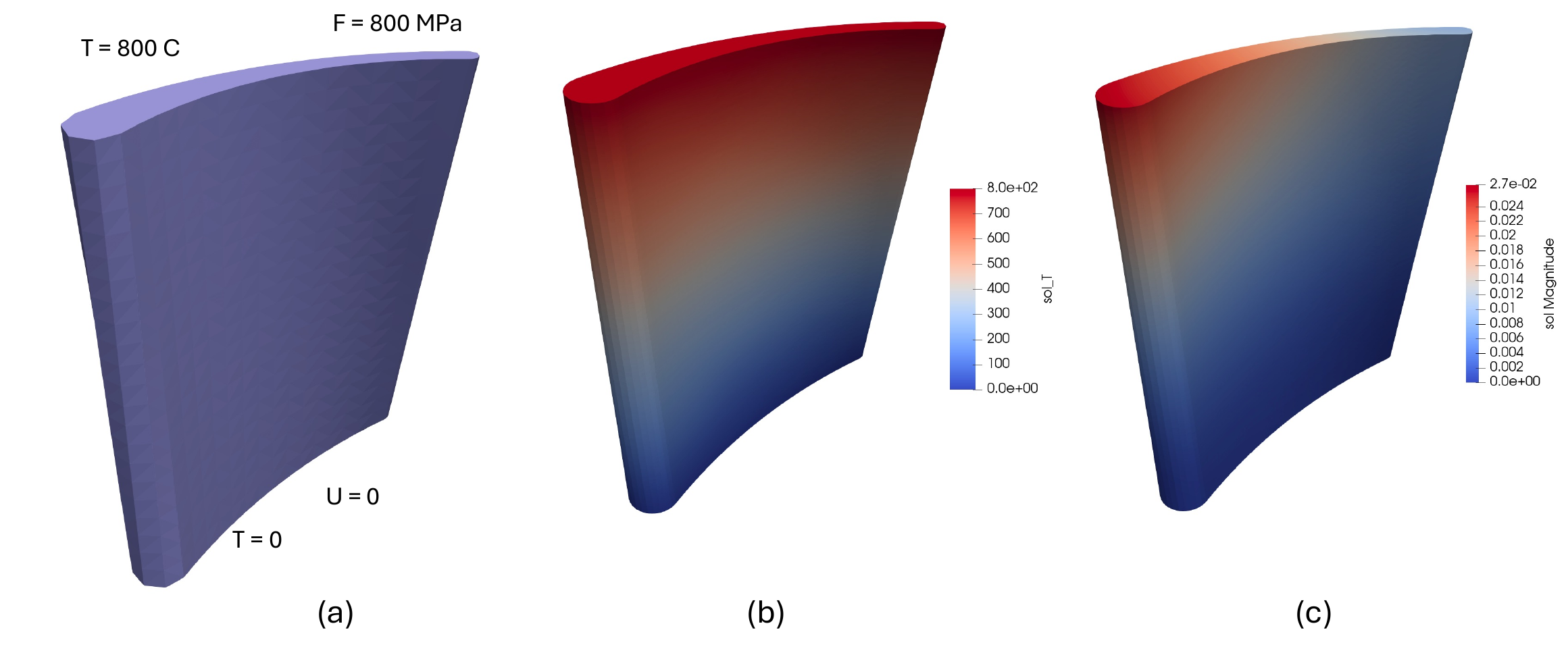}
 		\caption{(a) The geometry of the blade and the imposed boundary condition. (b) The resulting Temperature distribution (c) and magnitude of displacement.}
 		\label{fig:blade_bc_temp_disp}
	\end{center}
\end{figure}

This example considers the optimization of a turbine blade under thermal-mechanical loading \cite{Knapik2025}. Specifically, we examine a simplified scenario aimed at minimizing the thermally induced structural compliance.The design domain along with the imposed boundary conditions are shown in \Cref{fig:blade_bc_temp_disp}(a). The blade is optimized for a maximum mass of 3000 (39.5\% of the mass of a pure ${\rho}_1$ design). 
The idealized material composition presented in \Cref{fig:material_properties} is used to populate the design region. The resulting  temperature and displacement fields for $\bar{B} = 0.1$ is illustrated in \Cref{fig:blade_bc_temp_disp}(b) and (c) respectively.

\begin{figure}[h]
 	\begin{center}
		\includegraphics[scale=0.3,trim={0 0 0 0},clip]{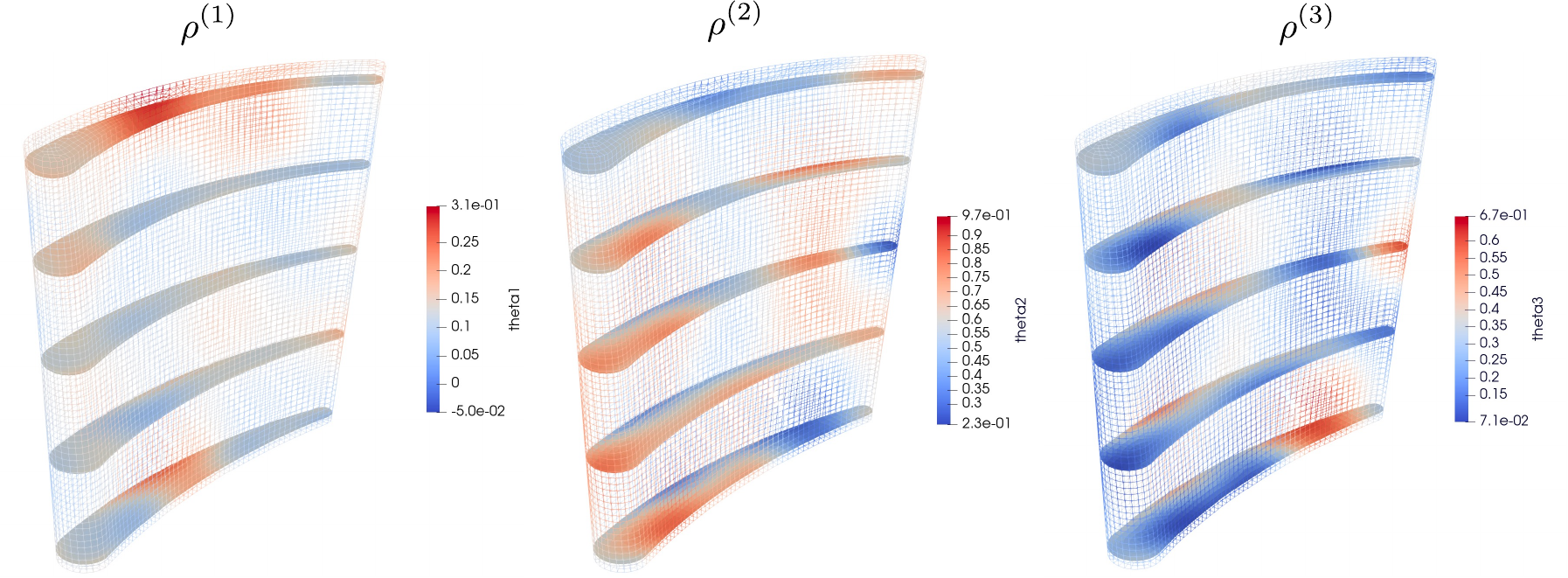}
 		\caption{The distribution of the three components at various slices for optimization with $\bar{B} = 0.1$.}
 		\label{fig:blade_composition_dist}
	\end{center}
\end{figure}

\begin{figure}[h]
 	\begin{center}
		\includegraphics[scale=0.5,trim={0 0 0 0},clip]{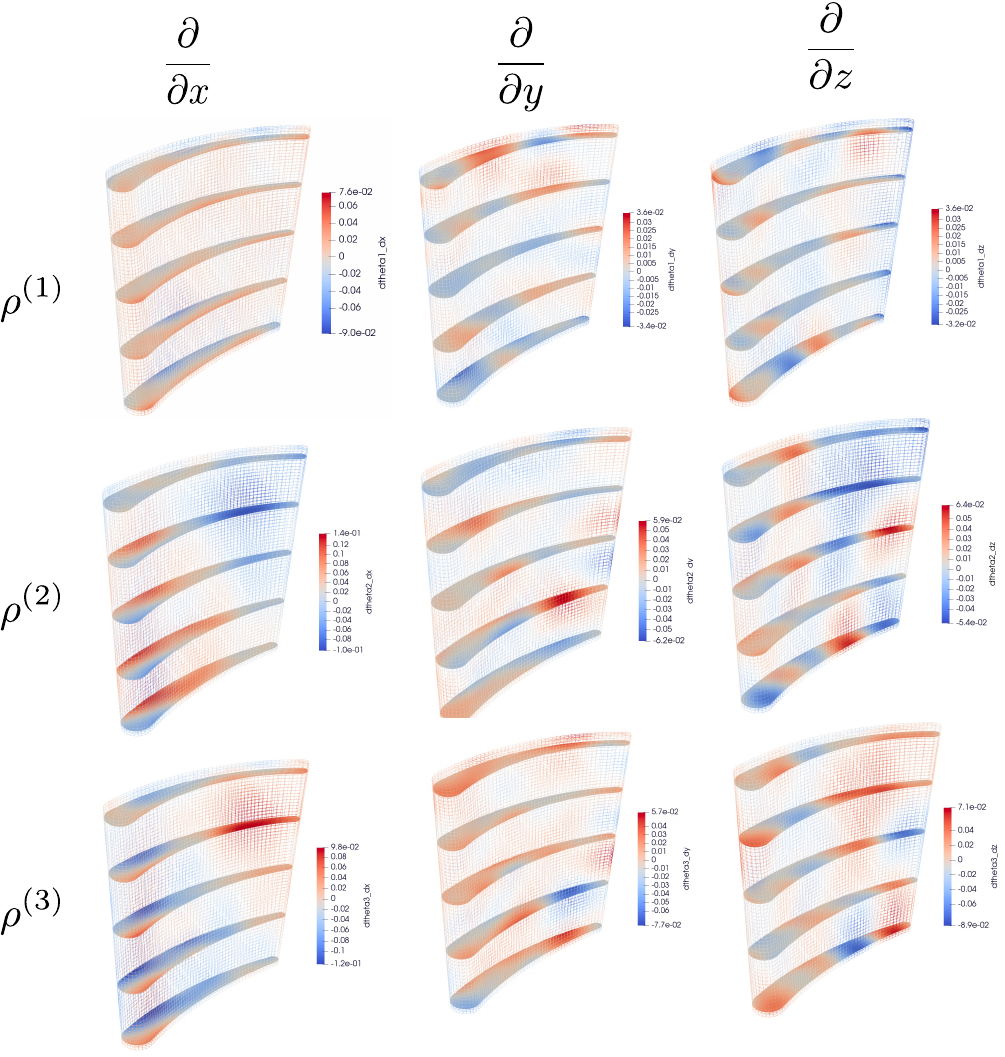}
 		\caption{The gradient of the three components along the X,Y and Z axis. }
 		\label{fig:blade_grad_compositoon}
	\end{center}
\end{figure}

\begin{figure}[h]
 	\begin{center}
		\includegraphics[scale=0.5,trim={0 0 0 0},clip]{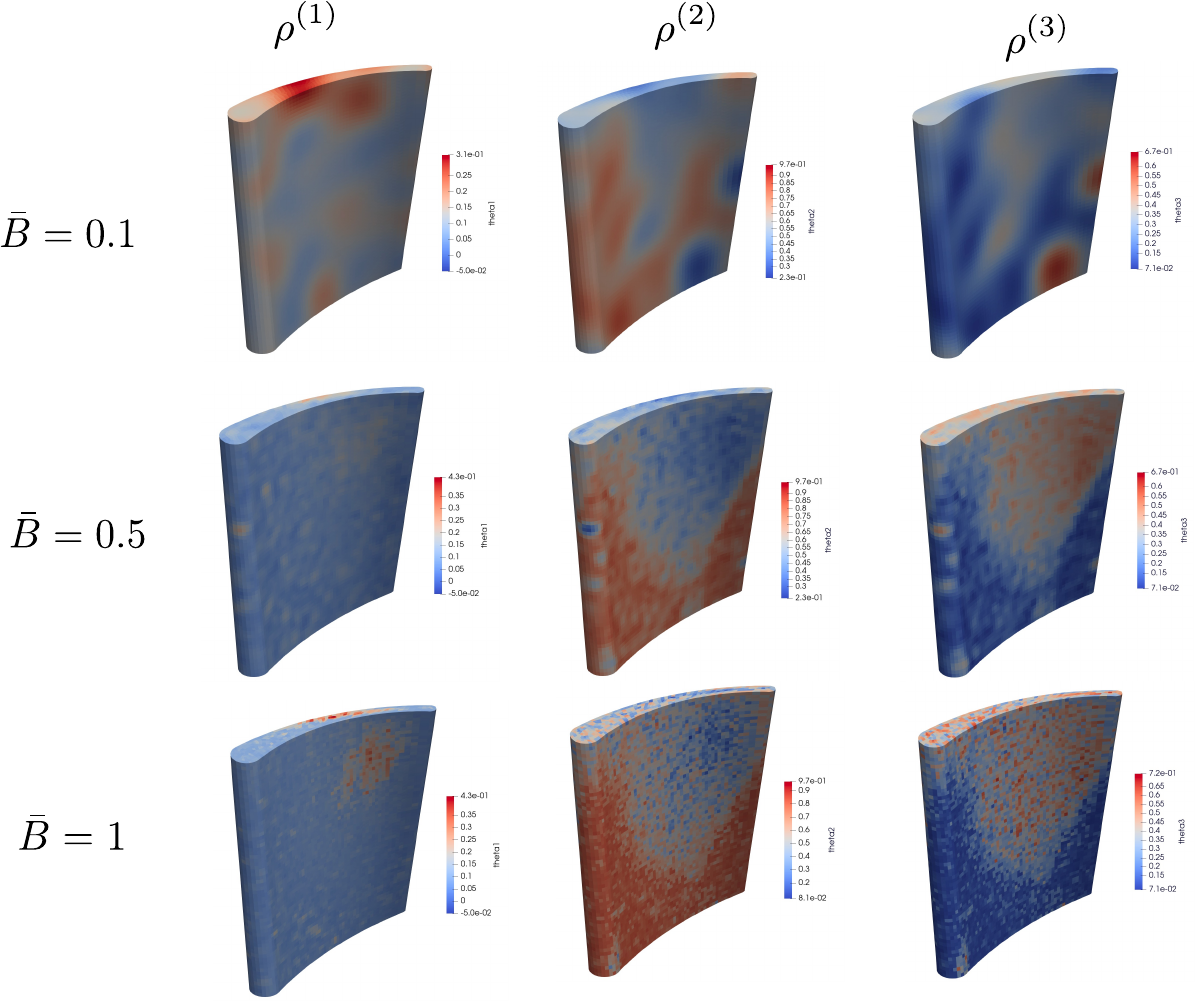}
 		\caption{The optimized compositions for various gradation limits. }
    \label{fig:blade_compositon_bandwidth}
	\end{center}
\end{figure}

 Furthermore, the composition distribution for a bandlimit of $\bar{B} = 0.1$ is shown in \Cref{fig:blade_composition_dist}. We also illustrate the gradient of the composition of the three components along the X,Y, and Z axis in \Cref{fig:blade_grad_compositoon}. We observe that the gradations once again are confined to the imposed gradation limit. Finally, we vary the gradation limit $\bar{B} = \{0.1, 0.5, 1\}$  along the X,Y and Z axis respectively and illustrate the obtained composition distribution in  \Cref{fig:blade_compositon_bandwidth}. As expected, we see an increased gradation as we increase the allowed bandwidth. Furthermore, the relative compliances are 1, 0.76 and 0.62 respectively; indicating an improvement in performance as we allow for increased gradation. Although this is a simple illustrative example, the use of CGA has been shown to be critical in reducing the amount of critical material used. This experiment validates that the proposed method is applicable for constraining material gradation in real-world applications.

\section{Conclusion}
\label{sec:conclusion}

This paper presented a topology optimization (TO) framework for designing gradation-limited structures using compositionally graded alloys (CGAs). By optimizing the spatial distribution of CGA composition, the framework enables the design of stronger, lighter, and more cost-effective components. The framework addresses the critical challenge of constraining the maximum spatial gradation of composition, a key limitation in the additive manufacturing of CGAs. We achieve this using a band-limited coordinate neural network that maps spatial coordinates to component fractions. The network's maximum bandwidth can be tuned to implicitly enforce a band-limited design, which, by Bernstein's inequality, limits the gradation in the final designs.

A key advantage of our method over conventional element-based TO is the use of a neural network as a continuous and mesh-independent design representation. Traditional methods struggle with accurately modeling the continuous nature of CGAs and often rely on inaccurate finite-difference calculations for spatial gradients. Furthermore, their design resolution is tied to the analysis mesh, restricting design complexity. Our neural network approach overcomes these issues. It implicitly enforces gradation control by regulating the network's bandwidth, which eliminates the need for numerous explicit constraints. This leads to a more robust and efficient optimization process. The framework is also end-to-end differentiable, streamlining gradient computations through automatic differentiation.

We demonstrated the method's effectiveness through several thermo-elastic TO examples, which produced optimized CGA designs adhering to prescribed gradation limits. The framework is adaptable to various manufacturing constraints, such as different gradation limits along different axes. Other benefits include the ability to extract high-resolution designs and an end-to-end differentiable process.

While the current framework is effective, it assumes an idealized relationship between composition and material properties. Future work will extend the material model to include dependencies on both temperature and composition and will investigate practical gradation constraints derived from manufacturing limits or thermal stress. We also plan to incorporate other manufacturing considerations into the optimization, such as limiting the phase fraction of certain constituents, and considering the interaction between process, material, and geometry. To ensure practical designs, we will add constraints to avoid undesirable phase compositions. A critical next step is the experimental validation of these designs by fabricating the CGA components and testing their mechanical performance. Finally, the utility of CGAs is significant in fields like aerospace, where factors such as oxidation, creep, and corrosion are critical. Future efforts will extend the framework to include these factors.

\section*{Acknowledgments}
This work was supported by the Defense Advanced Research Projects Agency (DARPA) Multiobjective Engineering and Testing of Alloy Structures (METALS) program project titled “RADICAL: Rapid Array DImple based Co-design of gradient materiaL and geometry” under cooperative agreement No. HR0011-24-2-0302. This work was also supported by the National Science Foundation grant 2219489 to WC and the Department of Defense Vannevar Bush Faculty Fellowship, N00014-19-1-2642, to JC. The authors extend their thanks to Rujing Zha and Rowan Rolark for their assistance in the 3D printing process.

\section*{Compliance with ethical standards}
The authors declare that they have no conflict of interest.

\section*{Replication of Results}
The implementation will be made available upon reasonable request.

\bibliographystyle{unsrt}  
\bibliography{references}  

\end{document}